\documentclass[conference]{IEEEtran}
\IEEEoverridecommandlockouts
\usepackage{cite}
\usepackage[colorlinks=true, linkcolor=black, citecolor=black, urlcolor=blue]{hyperref}
\usepackage{amsmath,amssymb,amsfonts}
\usepackage{graphicx}
\usepackage{textcomp}
\usepackage{xcolor}
\usepackage{tabularx}
\usepackage{cite}
\usepackage{array}
\usepackage{multicol}
\usepackage{multirow}
\usepackage{makecell}
\usepackage{booktabs} 
\usepackage{algorithmic}
\usepackage[linesnumbered,ruled,vlined]{algorithm2e}
\usepackage{csquotes}
\usepackage{float}   
\usepackage{enumitem}        

\def\BibTeX{{\rm B\kern-.05em{\sc i\kern-.025em b}\kern-.08em
    T\kern-.1667em\lower.7ex\hbox{E}\kern-.125emX}}
\begin{document}

\title{Heterogeneous Multi-Agent Modeling for Measurement and Network Analysis of the Data Service Market
\\
\thanks{* Corresponding author: Lizhen Cui; e-mail: clz@sdu.edu.cn;}
}

\author{\IEEEauthorblockN{
Deyu Zhou\textsuperscript{1,2},  
Yuwei Guo\textsuperscript{3},
Xudong Lu\textsuperscript{1,2}, 
Linhao Zhang\textsuperscript{1,2},
Wei Guo\textsuperscript{1,2},
Lizhen Cui\textsuperscript{1,2*}} \\

\IEEEauthorblockA{
\textit{\textsuperscript{1} School of Software, Shandong University, Jinan, China} \\  
\textit{\textsuperscript{2} Joint SDU-NTU Centre for Artificial Intelligence Research (C-FAIR), Shandong University, Jinan, China} \\ 
\textit{\textsuperscript{3} College of Intelligence and Computing, Tianjin University, Tianjin, China} \\
} \\
Email: zhoudeyu@mail.sdu.edu.cn, 17860130125@163.com, dongxul@sdu.edu.cn, \\ linhaoz2022@163.com, guowei@sdu.edu.cn, clz@sdu.edu.cn}

\maketitle

\begin{abstract}
With the increasing complexity of collaboration among various social entities and user demands, the factors affecting the stable development of the data service market are also growing. These factors include the widespread dissemination of information enhancing subjective consciousness, the continuous improvement in intelligence, and the complexification of structural relationships. To achieve effective governance and regulation of the data service market, it is crucial to conduct simulation experiments before making regulatory decisions. However, current research and analysis of the data service market primarily focus on data-level performance, proving inadequate when it comes to measurement and analysis of multiple heterogeneous entities and the integration of various social elements within the data service market. Based on this, this paper innovatively proposes a data service market measurement and network analysis method based on heterogeneous multi-agent modeling. By introducing the service ecosystem theory, we clarify the participants and external factors of the data service market and conduct utility measurements for three-level entities based on value creation. Furthermore, an analytical methodology is devised to precisely assess the influence of heterogeneous networks on utility. Finally, the paper verifies the effectiveness of the proposed method through the analysis of experimental results.
\end{abstract}


\section{Introduction}

The data service market is a supply-demand matching system in the modern service industry, which is driven by data and provides digital customized services for market demands through the division of labor and collaborative operation of multiple individuals. As the social structure is redefined under the logic of services \cite{gronroos2014service}, the data service market is gradually transforming into a service ecosystem in which numerous intelligent service entities (such as humans, institutions, software robots, etc.) operate collaboratively \cite{driessen2022data}. It is no longer a single market of the traditional data industry, but a multi-factor integrated market that spans industries and meets various demands\cite{liang2018survey}. 

\begin{figure*}[ht]
  \centering
  \includegraphics[width=\textwidth]{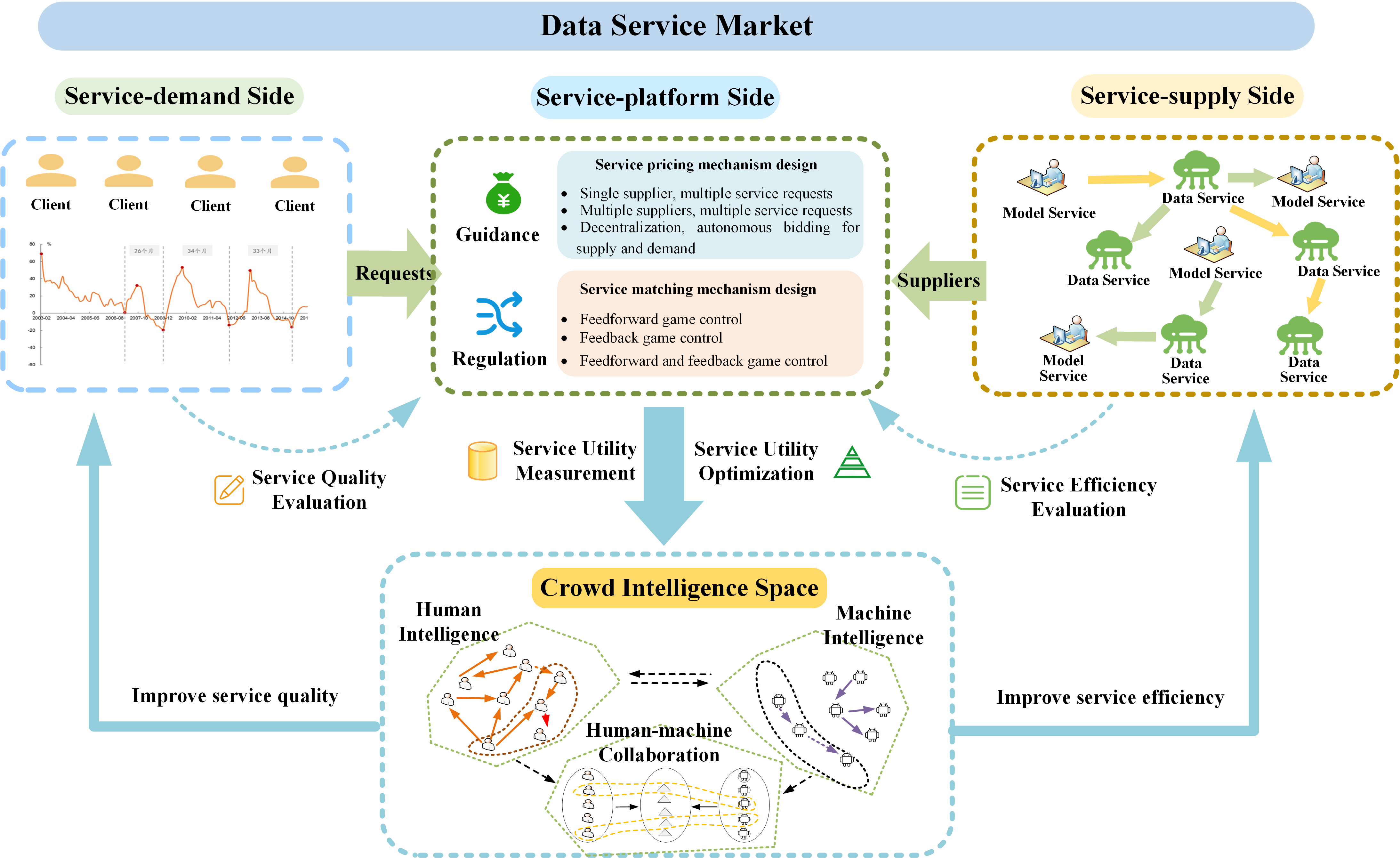}
  \caption{The Operational Structure of Data Service Market}
 \label{Pic1}
\end{figure*}

The data service market includes the demand side, the platform side, and the supply side. Their logical relationships are shown in Fig.\ref{Pic1}. Customers on the demand side make service requests through passive guidance or active exploration. Model and data service providers on the service supply side offer services \cite{lu2021computational}. The service platform side designs pricing mechanisms and matches requests with supplies. Various intelligent service entities have formed complex inter-relationships through long-term competition and cooperation, and exhibit adaptive behaviors to the environment.By measuring and optimizing service utility, the process of cooperation between human agents and robot agents in the collective intelligence space is regulated, and governance strategies are adjusted accordingly, ultimately improving service quality and utility \cite{centola2022network,cui2024ai}. 

On one hand, the data service market is a bottom-up, self-organizing, and self-growing complex system, where the overall state of the system depends on the spontaneous interactive behaviors among different service entities \cite{xue2016computational}. On the other hand, the data service market is also a product of top-down planning, where managers can use patterns obtained from various analytical methods to alter the design planning and evolution path of the data service market \cite{graef2015market}. Due to the inherently low visibility of the service process, the visibility of risks can easily be obscured at various stages, leading to their accumulation until they eventually erupt. Therefore, how to achieve effective monitoring and analysis of the data service market has become a crucial issue for its sustainable and healthy development \cite{kumar2013data}.

As a complex and uncertain social system, the regulation of the data service market exhibits dynamic and diversified trends, which impose higher demands on research methods \cite{feng2020review}. Researchers mainly analyze service ecosystems through data-driven statistical analysis method and agent-based modeling (ABM) \cite{xue2023chatgpt, railsback2019agent}. However, current research and analysis of the data service market primarily focus on data-level performance, proving inadequate when it comes to measure and analyze of multiple heterogeneous entities and the integration of various social elements within the data service market\cite{visnjic2018path}.

Based on this, in this paper, we innovatively propose a measurement and network analysis method for the data service market based on heterogeneous multi-agent modeling from the perspective of service ecosystem. By modeling the heterogeneous entities and their relationships, we define a service utility measurement model for the data service market \cite{xue2023computational}. Furthermore, a method for analyzing the impact of heterogeneous networks on utility is established.

The rest of this paper is organized as follows: Section 1 presents the research background and motivation; Section 2 presents the measurement and network analysis methods for data service market; Section 3 gives the construction and design of experiments system about data service market, and analyzes the experimental results; Section 4 concludes the paper.

\section{Background and Motivation}
The influencing factors in the evolution process of the data service market are complex and difficult to identify. They are not only affected by human free will and subjective initiative but can also change their evolutionary path due to occasional events. Therefore, it is necessary to study its evolutionary analysis methods. Currently, researchers commonly use the following methods to analyze the data services market:

\textbf{Data-driven Statistical Analysis}: Big data lays more emphasis on application. It is a method of predicting phenomena by collecting large amounts of data. Its drawback is that although it can make predictions, it is difficult to explain the predicted phenomena. This method continuously monitors system operation data to detect potential faults or anomalies. For example, statistical model-based anomaly detection techniques construct a data model from all the data and identify outliers as objects that do not fit the model perfectly. This method is very challenging when studying complex systems that are difficult to represent with mathematical models \cite{markovsky2021behavioral}. Based on a systematic review of data market research literature, Abbas et al. believe that the current understanding of the data service market abroad is still unclear\cite{abbas2021business}. Technical research on the data market mainly focuses on the construction of data trading platforms, especially the application of blockchain. Mahajan et al. proposed a blockchain-based data market development method from the perspective of enhancing the capabilities of data users\cite{mahajan2022data}.

\textbf{Agent-based Modeling (ABM)}: Compared with data-driven methods, ABM focuses more on theory. Its function is to extract theories and put forward testable and falsifiable hypotheses through computer simulation. It can not only propose predictive hypotheses about phenomena but also explain the phenomena and their causes through theory\cite{peng2023computational,zhou2025federated}. Agent-based modeling is very suitable for describing the dynamic evolution process of complex systems. However, there are some limitations in using ABM methods for service ecosystem simulation modeling, such as the lack of research on heterogeneous intelligent agent systems and the absence of comprehensive metrics for system measurement during ABM applications \cite{xue2022research}. 

In the traditional industrial era, enterprises embedded value into products during the production process and realized value transfer and appreciation through market transactions\cite{huang2020value,sjodin2020value}. Vargo et al. proposed the concept of service ecosystem, believing that the service ecosystem is a self-regulating and relatively independent system formed by loosely coupled participants connected by institutions and value co-creation mechanisms\cite{vargo2010repeat,xueandYu2024computational}. Besides, with the development of the internet technologies, social networks have become an indispensable part of human daily life. People share information, establish connections, and obtain services through social networks. Researchers have begun to explore how to use graph theory and complex network analysis methods to model the relationships between service providers and customers in social networks\cite{milroy2013social}. They describe entities as nodes and study their interactions through edges in the social network~\cite{zhou2024hierarchical}. These methods have revealed some characteristics and patterns of complex adaptive systems in social networks.

Current research methods prioritize prediction over explanation, making it difficult to reveal the underlying mechanisms of the data service market's evolution and lacking a systematic characterization of heterogeneous intelligent agents within the service ecosystem~\cite{brunner2024towards,kumar2016research,xue2024computational}.
To address the limitations of existing analysis methods in studying the data service market, this paper proposes a data service market measurement and network analysis method based on heterogeneous multi-agent modeling. From the perspective of service ecosystem, this method clarifies the participating entities, value creation patterns, and external elements of the data service market. On this basis, a service utility measurement model and a heterogeneous agent collaboration model based on complex networks are defined.

\section{Measurement and Network Analysis Method for the Data Service Market}
This section addresses the challenges of complexity representation and system complexity assessment of heterogeneous entities in the data service market by modeling and measuring heterogeneous multi-level agents. Finally, a method for analyzing the impact of heterogeneous networks based on service utility is designed\cite{xue2019analysis}, Table \ref{tab:key_indicators} summarizes the important indicators.

\begin{table}[H]
\centering
\caption{Key indicators used in measurement and network analysis for the data service market.}
\label{tab:key_indicators}
\begin{tabular}{p{1.2cm} p{6cm}}
\hline
\textbf{Indicator} & \textbf{Explanation} \\
\hline
$N_{\Delta t}(t)$ &
The number of order arrivals within the interval $[t,t+\Delta t]$. \\
$U^{\mathrm{dem}}(\Delta t)$ &
The utility on the demand side at time $t$. \\
$D^{\mathrm{op}}(t)$ &
The platform’s decision objective. \\
$U^{\mathrm{collab}}(o)$ & The collaboration utility for an order $o$ executed by an agent set. \\
$U_r$ &
The individual utility of robot $r$. \\
$U_p$ & The individual utility of business agent $p$. \\
$U_G$ & The organizational utility of $G$. \\
$U_{\text{sys}}$ &
The system utility at the platform level. \\
\hline
\end{tabular}
\end{table}

\subsection{Structure-constrained Heterogeneous Agent Modeling in Data Service Market}
From the perspective of service ecosystems, the evolution of the data service market is based on supply and demand matching. The system includes the service demand side, service supply side, and service operation side\cite{syrbe2017ecosystem,wei2017integrating,xue2016computational}. In the data service market, the purpose of measurement and analysis is to better govern the service system on the supply side. Therefore, the nodes on the demand side are regarded as the external market environment and are represented by the demand fluctuation sequence, while the agents in the service system on the supply side are the focus of modeling. The modeling of data service market from these three aspects is as follows:

\begin{itemize}
\item {\textbf{Demand-side Modeling}}
\end{itemize}

Demands are manifested as a stream of user orders arriving over time. To characterize demand uncertainty and heterogeneity, we consider the inter-arrival pattern of orders, the order volume within a time window, and the composition and complexity of orders, including order types, processing difficulty, and the proportion of different categories.

We model order arrivals using queuing theory and assume that the number of orders arriving within a short time window follows a Poisson process. To regulate the workload injected into the system, we introduce a user-adjustable \emph{order arrival rate} that controls the expected number of arrivals per unit time:
\begin{equation}
N_{\Delta t}(t) \sim \mathrm{Poisson}\left(\lambda(t)\Delta t\right),
\label{eq:poisson_arrival}
\end{equation}
where, $\lambda(t)$ is the order arrival rate at time $t$, and $\Delta t$ is the window length. The function $\lambda(t)$ can be specified as a constant (stationary demand) or a time-dependent profile (non-stationary demand) to capture peak hours, bursts, or periodic fluctuations.

 Given the arrivals, we generate order attributes via Monte Carlo sampling to reflect heterogeneity in order types and processing requirements. Specifically, for each arriving order, we sample its type, difficulty, and other features from pre-defined distributions, and the proportions of different types are controlled by scenario parameters.
When historical order data are available, we estimate the empirical distributions (or fit parametric forms) from real observations and then sample accordingly, enabling data-driven simulation of order fluctuations and composition. When real data are unavailable, we adopt synthetic yet controllable distributions by specifying a fluctuation function for $\lambda(t)$ (e.g., sinusoidal or piecewise functions) and selecting plausible priors for order-type proportions and difficulty, which together produce scenario-consistent demand streams for computational experiments.


The duration and quality of order completion can be used to represent how well the system performs the task, while the time and cost spent by the system to complete all incoming orders can measure the overall capability. We define demand side utility as a \enquote{quality–time–cost} trade-off:
\begin{equation}
U^{\mathrm{dem}}(\Delta t)
= \mathbb{E}\!\left[ q(o) - \lambda_T T(o) - \lambda_C C(o) \right]
\label{eq:dem_utility}
\end{equation}
where, $o$ denotes an incoming order during $\Delta t$; 
$q(o)$ represents the service quality obtained from completing order $o$;
$T(o)$ is the processing time of order $o$;
$C(o)$ denotes the processing cost;
$\lambda_T$ and $\lambda_C$ are penalty coefficients associated with time and cost, respectively;
$\mathbb{E}[\cdot]$ takes expectation over order arrivals.

\begin{itemize}
\item {\textbf{Operation-side Modeling}}
\end{itemize}

The operation side refers to the system management platform agent, which coordinates and controls the entire system, receives orders from external sources, allocates orders to different institutions for processing, and calculates total cost and revenue to evaluate overall utility. The platform’s decision objective is modeled as:
\begin{equation}
D^{\mathrm{op}}(t) = R(t) - \eta_{1}\,\overline{T}(t) - \eta_{2}\,\Omega(t),
\label{eq:platform_utility}
\end{equation}
where, $R(t)$ is the total revenue (or value) obtained from processed orders; 
$\overline{T}(t)$ denotes the average order processing time; 
$\Omega(t)$ is a load imbalance term across institutions (e.g., variance of 
occupancy); and $\eta_{1}, \eta_{2}$ are penalty coefficients.

The platform assigns orders to institutions by prioritizing allocation according to institutional occupancy levels, while jointly considering order complexity as well as robot and business agent utilization. Accordingly, this decision logic is formulated as a structure-constrained assignment problem:
\begin{equation}
\begin{aligned}
x_{o,i} 
& = \arg\max_{i \in \mathcal{I}}
\left(
\hat{u}_{o,i}
- \kappa_{r}\,\rho_{i}^{r}(t)
- \kappa_{e}\,\rho_{i}^{e}(t)
\right), \\
&  \text{s.t.}\quad
\sum_{i \in \mathcal{I}} x_{o,i} = 1 .
\end{aligned}
\label{eq:assignment}
\end{equation}
where, $\mathcal{I}$ denotes the set of institutions; 
$x_{o,i} \in \{0,1\}$ indicates whether order $o$ is assigned to institution $i$; 
$\hat{u}_{o,i}$ represents the estimated utility if institution $i$ processes 
order $o$ (reflecting order difficulty, expected processing time, and cost); 
$\rho_{i}^{r}(t)$ and $\rho_{i}^{e}(t)$ denote the robot and business agent occupancy 
rates, respectively; and $\kappa_{r}, \kappa_{e}$ are penalty weights.

\begin{itemize}
\item {\textbf{Supply-side Modeling}}
\end{itemize}
The supply side is the main body of the system, forming a multi-organization and multi-agent system where different organizations collaborate to fulfill orders. 
The supply side includes secondary entities (organizations) and primary entities within organizations (high intelligence business agents and low-intelligence software robot agents). Organizational agents receive orders allocated by the platform; orders can be transferred between organizations. Within each organization, intelligent agents collaborate to handle system orders. High-intelligence agents handle higher-difficulty orders but consume more cost and time, while low-intelligence agents process quickly at lower cost but can only handle less difficult orders \cite{xue2021computational}. 
Embedding \emph{structure-constrained collaboration} , we define the collaboration utility of order $o$ executed by an agent set $\mathcal{A}_o$ as:
\begin{equation}
U^{\mathrm{collab}}(o)
=
\sum_{a \in \mathcal{A}_o} u_a(o)
-
\phi \, \lvert \mathcal{A}_o \rvert ,
\label{eq:collab_utility}
\end{equation}
where, $\mathcal{A}_o$ denotes the collaborating agent set; 
$u_a(o)$ is the individual utility contribution of agent $a$; 
$\lvert \mathcal{A}_o \rvert$ is the number of collaborators; 
and $\phi$ is the coordination overhead coefficient, capturing 
communication and coordination costs.
The collaboration must satisfy structural constraints induced by capability thresholds and availability:
\begin{equation}
\begin{aligned}
d(o) \le \theta_r 
&\;\Rightarrow\; a \in \mathcal{A}_o \cap \mathcal{R} \text{ is feasible}, \\
d(o) > \theta_r 
&\;\Rightarrow\; a \in \mathcal{A}_o \cap \mathcal{B} \text{ is required}, \\
\rho_a(t) < 1, 
&\;\forall a \in \mathcal{A}_o ,
\end{aligned}
\label{eq:structural_constraints}
\end{equation}
where, $d(o)$ denotes order difficulty; 
$\theta_r$ is the robot capability threshold; 
$\mathcal{R}$ and $\mathcal{B}$ denote the robot-agent set and business-agent set, 
respectively; 
and $\rho_a(t)$ is the occupancy rate of agent $a$ at time $t$ (which must be less than $1$ for availability).

\subsection{Utility Measurement Model of Data Service Market}
Heterogeneous agents create value through order processing.
To quantitatively evaluate system utility, an effective-value-based utility measurement model is introduced.
Different from coarse-grained evaluation limited to the organizational or platform level, system utility is decomposed into primary-entity utility, organizational utility, and platform-level utility, forming a consistent multi-level measurement framework \cite{byrne2016review}. The value created by the agents is a major source of system cost and revenue.

\begin{itemize}
\item {\textbf{Primary Entity: Individual Utility}}
\end{itemize}

Individual utility reflects the execution capability and efficiency of an agent during order processing. A higher individual utility indicates that collaborative orders can be completed in a shorter time, thereby reducing the actual delivery time of an order and improving order fulfillment quality and customer satisfaction. Software robot agents create effective value by completing orders while simultaneously incurring maintenance costs \cite{xue2021soa}. Accordingly, the individual utility of a robot is defined as:
\begin{equation}
U_r
=
\sum_{o \in \mathcal{O}_r}
\left(
\frac{R_h}{\sum A_h}
\cdot
O_{dv}(o)
\right)
-
C_r ,
\label{eq:robot_utility}
\end{equation}
where,
$\mathcal{O}_r$ is the set of orders processed by robot $r$;
$R_h$ represents the workload completed by robot $r$ in order $o$;
$\sum A_h$ denotes the total workload completed by all agents participating in order $o$;
$O_{dv}(o)$ represents the complexity or value of order $o$;
$C_r$ denotes the maintenance cost incurred by robot $r$ within a service cycle.
Business agents create effective value through order processing. Accordingly, the individual utility $U_p$ of a business agents is defined as \cite{yu2025unlocking}:
\begin{equation}
U_p
=
B
+
\sum_{o \in \mathcal{O}_p}
\left(
\frac{P_h}{\sum A_h}
\cdot
O_{dv}(o)
\cdot
O_{vr}
\right),
\label{eq:personnel_utility}
\end{equation}
where,
$B$ represents the base income of the business agent;
$\mathcal{O}_p$ denotes the set of orders processed by business agent $p$;
$P_h$ is the workload completed by business agent $p$ in order $o$;
$\sum A_h$ denotes the total workload of order $o$;
$O_{dv}(o)$ represents the value or complexity of order $o$;
$O_{vr}$ denotes the conversion coefficient that maps order value to income.

\begin{itemize}
\item {\textbf{Secondary Entity: Organizational Utility}}
\end{itemize}

Organizational utility characterizes the overall operational performance of an organization over a service cycle \cite{matthews2011assessing,mcadam2014role,xue2020value}. It is jointly determined by the total effective value created by internal agents and the organizational operating costs.
The created value of organization $G$ is defined as the total complexity-based value of the orders it processes:
\begin{equation}
CV_G = \sum_{o \in \mathcal{O}_G} O_{dv}(o),
\label{eq:org_created_value}
\end{equation}
where, $\mathcal{O}_G$ denotes orders set processed by organization $G$.
The organizational cost consists of robot maintenance costs and business agent compensation:
\begin{equation}
C_G = \sum_{r \in \mathcal{R}_G} C_r + \sum_{p \in \mathcal{P}_G} U_p ,
\label{eq:org_cost}
\end{equation}
where, $\mathcal{R}_G$ and $\mathcal{P}_G$ denote the sets of robots and business agents within organization $G$, respectively; $C_r$ represents the maintenance cost of robot $r$; and $U_p$ denotes the income of business agent $p$, which is regarded as an organizational expense \cite{zhou2022sle2}.
The organizational utility of organization $G$ is finally defined as:
\begin{equation}
U_G = CV_G \cdot G_{vr} - C_G ,
\label{eq:org_utility}
\end{equation}
where $G_{vr}$ is the organization value conversion coefficient.

\begin{itemize}
\item {\textbf{Tertiary Entity: System Utility}}
\end{itemize}

The platform is the operator of the service system, service's utility can help platform agent adjust system parameters. In this work, system utility is evaluated at the platform level, while agent-level and organizational-level utilities affect system utility implicitly through delivery efficiency and operational cost. Under the above mechanism, system utility $U_{\mathrm{sys}}$ at the platform level can be uniformly expressed as:
\begin{equation}
U_{\mathrm{sys}}
=
\sum_{o \in \mathcal{O}}
O_{dv}(o)\,
\frac{T_e(o)}
{T_{rly}(o)}
-
\sum_{G \in \mathcal{G}} C_G(E_G),
\label{eq:system_utility}
\end{equation}
where,
$\mathcal{O}$ denotes the set of all orders in the system;
$O_{dv}(o)$ represents the complexity or intrinsic value of order $o$;
$T_e(o)$ denotes the average processing time of historical orders with comparable complexity, serving as a reference baseline;
$T_{rly}(o)$ denotes the actual completion time of order $o$, which is a function of the individual utility set $\{U_a\}_{a \in \mathcal{A}_o}$ and the organizational utility set $\{E_G\}_{G \in \mathcal{G}}$, reflecting the combined influence of individual execution efficiency and organizational coordination on delivery performance;
$\mathcal{G}$ denotes the set of organizations in the system;
$C_G(E_G)$ denotes the operational cost of organization $G$, which is modeled as a function of organizational utility, reflecting resource consumption, maintenance overhead, and collaboration costs.

In implementation, the realized completion time $T_{rly}(o) $ is computed by a workload–capacity–congestion mechanism, where individual utility affects effective processing speed and organizational utility affects the congestion multiplier. Specifically, the order workload is mapped from order difficulty, while the effective processing rate of each agent is determined by its baseline capability and individual utility \cite{van2024agent,shen2006agent,xue2018social}. In addition, collaboration size and average occupancy introduce multiplicative delay factors that capture coordination overhead and congestion effects.
\begin{equation}
T_{rly}(o)
=
\frac{W(o)}
{\sum_{a \in \mathcal{A}_o} s_a}
\cdot
\left( 1 + \lambda_{\rho}\,\bar{\rho}_{\mathcal{A}_o} \right)
\cdot
\left( 1 + \lambda_{c}\, (|\mathcal{A}_o| - 1) \right),
\label{eq:actual_completion_time}
\end{equation}
where, $W(o) = \omega_0 + \omega_1\, O_{dv}(o)$ denotes the workload of order $o$, which is mapped from its complexity. $\omega_0, \omega_1 > 0$ are scaling coefficients. $\mathcal{A}_o$ is the set of agents collaborating on order $o$, and $s_a$ represents the effective processing rate of agent $a$. $\bar{\rho}_{\mathcal{A}_o}$ denotes the average occupancy of the agents in $\mathcal{A}_o$, which reflects the workload pressure or resource utilization level of the collaborating agents when processing order $o$.
The effective processing rate of agent $a$ is defined as:
\begin{equation}
s_a = s_a^{0} \cdot \left( 1 + \lambda_u \, \tilde{U}_a \right),
\label{eq:agent_speed}
\end{equation}
where $s_a^{0}$ is the baseline processing rate of agent $a$ (which may differ 
between robots and human employees), and $\lambda_u \geq 0$ controls the influence 
of individual utility on processing speed. 
The normalized individual utility $\tilde{U}_a$ in Formula~\ref{eq:agent_speed} is given by:
\begin{equation}
\tilde{U}_a
=
\frac{U_a - \min(U)}
{\max(U) - \min(U) + \epsilon},
\label{eq:normalized_utility}
\end{equation}
where, $\epsilon$ is a small positive constant to avoid division by zero.
The average occupancy of the collaborating agent set $\bar{\rho}_{\mathcal{A}_o}$ in Formula~\ref{eq:actual_completion_time} is defined as:
\begin{equation}
\bar{\rho}_{\mathcal{A}_o}
=
\frac{1}{|\mathcal{A}_o|}
\sum_{a \in \mathcal{A}_o} \rho_a(t),
\label{eq:average_occupancy}
\end{equation}
where $\rho_a(t) \in [0,1]$ denotes the occupancy rate of agent $a$ at time $t$.
The coefficient $\lambda_{\rho} \geq 0$ captures congestion amplification due to 
high resource utilization, while $\lambda_c \geq 0$ represents the coordination 
overhead introduced by collaboration, increasing with the number of agents.

\subsection{A Structure-Constrained Network Analysis Method for Service Utility Evaluation}
This section develops a network-centric analysis method to research how heterogeneous interaction structures influence service utility in data service markets \cite{kapucu2013designing}.
We model the service ecosystem as an interaction network as follows:
\begin{equation}
\mathcal{W} = (\mathcal{N}, \mathcal{E}),
\label{G-N-E}
\end{equation}
where, nodes $\mathcal{N}$ represent service agents (human employees or robots), and edges $\mathcal{E}$ denote collaboration relationships established during service execution. The network structure serves as a structural constraint space, shaping feasible interaction patterns, coordination costs, and information flow efficiency.

Let $N=|\mathcal{N}|$ be the total number of agents. The interaction structure at time $t$ is characterized by
an adjacency matrix
\begin{equation}
\mathbf{A}_t = \big[a_{ij}(t)\big] \in \{0,1\}^{N\times N},
\end{equation}
where, $a_{ij}(t)=1$ indicates an active collaboration between Agent $i$ and Agent $j$.
The degree of an agent $k_i(t)$ measures collaboration capacity, reflecting ability to share workload, exchange information, and participate in joint service execution.
\begin{equation}
k_i(t) = \sum_{j=1}^{N} a_{ij}(t),
\end{equation}

At the system level, structural statistics such as average degree $\bar{k}$, clustering coefficient,
and path length collectively determine the interaction efficiency and coordination overhead
of the service network.
For the data service market, collaboration is inherently dynamic and history-dependent. We
quantify the cooperation intensity $P_{ij}$ between agents $i$ and $j$ as
\begin{equation}
P_{ij}=\frac{N_{ij}}{\frac{1}{T}\sum_{t=1}^{T} N_{ij}^{(t)}},
\label{eq:coop_intensity}
\end{equation}
where, $N_{ij}$ denotes the number of completed service interactions between the agent pair, and
$T$ is the total number of repeated runs.
This formulation captures long-term collaboration stability, serving as a weighted
interpretation of edges in $\mathcal{G}$, and bridges the network structure with realized service outcomes.

Under structural constraints imposed by the interaction network $\mathcal{W}$, the platform allocates a task set $\mathcal{T}$ to agents and produces a collaboration outcome
\begin{equation}
\mathbf{x}=\mathcal{A}(\mathcal{T},\mathcal{W}), \qquad 
\mathbf{x}=\{x_{it}\},
\end{equation}
where $x_{it}$ denotes the assignment outcome of agent $i$ on task $t$.
For each agent $i$, the effective service capability is not a constant, but is modulated by the network:
\begin{equation}
\tilde{s}_i = s_i \cdot \phi_i(\mathcal{W}),
\qquad
\phi_i(\mathcal{W})=\sum_{j} a_{ij} P_{ij},
\end{equation}
where $s_i$ is the baseline capability, $\phi_i(\mathcal{W})$ is a network-induced collaboration amplification factor,
$a_{ij}$ indicates whether $i$ can interact with $j$ in $\mathcal{W}$, and $P_{ij}$ denotes the interaction (or collaboration) strength/probability.
Thus, $\mathcal{W}$ directly shapes ``who can do what'' and ``to what extent'' in the mechanism.

\section{Experimental System of the Data Service Market}
We conducted extensive experiments to address a series of research questions regarding the method proposed in this paper \cite{xue2024computational}: 
It is mainly used to address a series of research questions:
\begin{itemize}[topsep=1pt,leftmargin=10pt,parsep=0pt]
    \item \textbf{RQ1:} Can the proposed service utility measurement method evaluate the data service market more stably and effectively?
    \item \textbf{RQ2:} How do collaboration patterns under different network structures influence the service utility of the data service market?
\end{itemize}

\subsection{Experimental System Construction}
The data service market is focused on the digital government system in our work, primarily describing the interaction behaviors and processes between service entities and the relationship between service entities and public demands. Citizens can use digital platforms to query government information, apply for administrative permits, or submit feedback. The government community processes and responds to these requests via the digital platform and provides corresponding services based on rules and processes. By modeling the operational mechanisms, processes, and relationships among participants in the digital government system, guidance can be provided for designing and improving digital government systems to enhance citizen engagement, service quality, and government utility.

\begin{figure}[htbp]
\centerline{\includegraphics[width=0.99\linewidth]{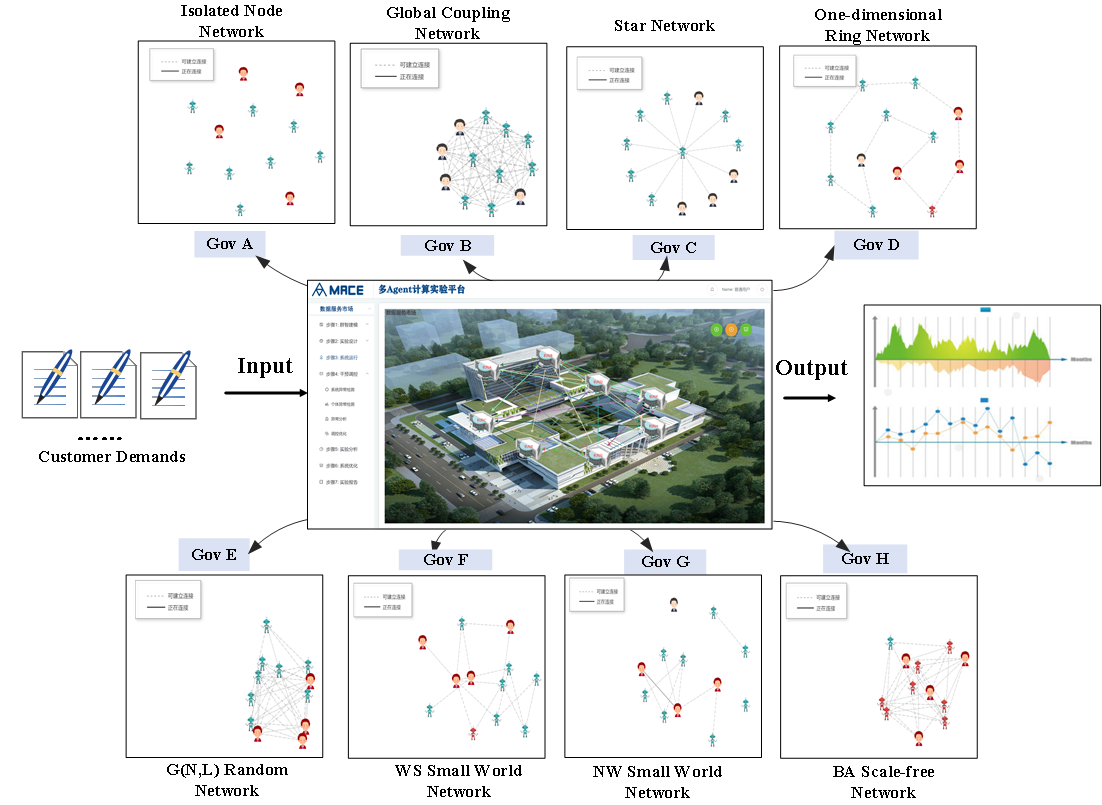}}
\caption{The Experimental Architecture of Data Service Market}
\vskip 10.1pt%
\label{Pic2}
\end{figure}

As shown in Fig.\ref{Pic2}, the system involves five major components: \enquote{Orders}, \enquote{Platform}, \enquote{Government Agencies}, \enquote{Salesman Agents}, and \enquote{Robots}. Each component needs to be designed as a distinct entity class. The system currently includes 8 government Agencies (i.e.,Institutions), each institution builds a collaborative structure corresponding to each network model, labeled A through H. Agency A can only process orders of type A, and so on. These institutions collaborate to complete customer orders. Each institution initially has 12 agents: 4 salesman agents (i.e., P1,P2,P3 and P4) and 8 robots. They work together to process system orders. Salesman agents can handle more difficult orders but incur costs and have slower processing speeds, while robots process orders quickly but can only handle less difficult orders.
Orders originate from outside the system and come in eight basic types: A, B, C, D, E, F, G, and H, as well as their random combinations. For example, BA type orders are equivalent to AB type orders, and so forth. Additionally, each order type has different levels of complexity. The complexity of an order indirectly represents its value. 

Different organizational and market conditions give rise to heterogeneous interaction topologies, which are modeled as structural priors:
\textbf{Isolated Network:} Agents operate independently ($k_i=0$), minimizing coordination cost but severely limiting collaborative capacity.
\textbf{Globally Coupled Network:} Fully connected structure enabling maximal collaboration, at the expense of high communication and coordination overhead.
\textbf{Star Network:} A centralized topology where a hub agent mediates interactions, introducing structural asymmetry and potential bottlenecks.
\textbf{One-Dimensional Ring:} Agents form a ring network, where each Agent can only form cooperative relationships with its two adjacent Agents.
\textbf{G(N,L) Random Network:} There are $N$ agent nodes are connected by randomly placed links. Where, $0\le\ L\le\frac{N\left(N-1\right)}{2}$, initially $L$ is set to $N(N-1)/8$.
\textbf{WS Small-World Network:} Consider a nearest-neighbor coupled network with $N$ nodes arranged in a ring, where each node is connected to $K/2$ nodes on its left and right (K is even). The parameters satisfy $N\gg K\gg lnN \gg1$. Then, with probability $p$, each edge in the network is randomly reconnected, i.e., one endpoint of the edge remains unchanged while the other endpoint is replaced with a randomly selected node in the network. The small-world network specifies that there should be no multiple edges or self-loops.
\textbf{NW Small-World Network:} Similarly, in a K-nearest neighbor network, an edge is added between a randomly selected pair of agent nodes with probability $p$. This addition of edges also ensures that there are no multiple edges or self-loops in the network.
\textbf{BA Scale-Free Network:} Starting from a network with $m_0$ nodes, each time a new node is introduced, it connects to $m$ existing nodes, where $m\le\ m_0$. The probability $\Pi_i$ that the new node connects to an existing node $v_i$ is proportional to the degree $k_i$ of node $v_i$ , i.e.,  $\Pi_i= \frac {k_i}{\sum_{j} k_j}$. After $t$ steps, this algorithm produces a network with $ N=\ t\ +\ m_0$ nodes and $m_t$ edges. Initially, the network is set to have two nodes, and each new node is connected to two existing nodes according to the preferential attachment mechanism.

\subsection{Parameter Settings}

\begin{figure}[htbp]
  \centering
  \includegraphics[width=0.9\linewidth]{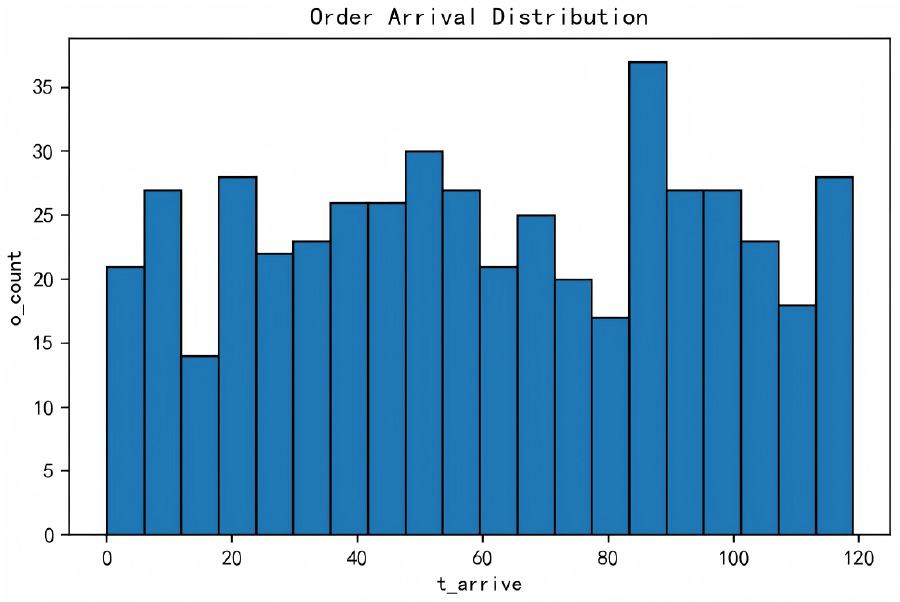}
  \caption{The Order Arrival Distribution}
  \label{fig:Order Arrival Distribution}
\end{figure}

By fitting the order distribution pattern on zbj.com, a digital service platform in reality, we take it as the order fluctuation data on the demand side of the data service market. Based on this, parameters about environment that can be designed for the system environment include\cite{yigitcanlar2024artificial}: \textbf{1)} Order Generation: The order quantity distribution based on real data is shown in Fig.~\ref{fig:Order Arrival Distribution}. \textbf{2)} Order Type: It is generated evenly corresponding to 8 types of institutions. \textbf{3)} Order Complexity: Using numeric Input Box, it allows input of positive integers only, with a minimum value of 1 and a maximum value of 3. \textbf{4)} Order Arrival Rate: It represents the average number of orders arriving per unit of time. Set between 0.5 and 10, allowing both integers and decimals. \textbf{5)} Cycle Duration: It allows input of positive integers only, with a minimum value of 60 and a maximum value of 240 (The default setting is 120). These parameters can be adjusted appropriately according to experimental requirements. 

For \textbf{RQ1}, we compare the proposed \emph{system utility}-based measurement ($U_{\text{sys}}$) with representative alternatives that are commonly adopted in service evaluation and information systems assessment. Specifically, we consider the following baselines: 
(i) \textbf{DeLone \& McLean (D\&M)} \cite{al2022empirical}: a system-success proxy constructed from observable variables (e.g., service quality, usage/intention, and net benefits) to approximate the classic information systems success model;
(ii) \textbf{QoS Utility (QoS)} \cite{mefgouda2024qos}: a traditional service-computing baseline that maps QoS attributes into a utility function and aggregates them with fixed weights, emphasizing performance-driven evaluation;
(iii) \textbf{Serv-qual proxy (SaL)}~\cite{seo2023role}: a service-quality baseline inspired by the expectation and perception gap, used to approximate perceived service quality in a measurable way. 
We report a unified set of outcome factors derived from the same simulation logs and standardize them to a common $[0,1]$ scale with consistent \emph{higher-is-better} semantics. For cost-type metrics we further reverse the normalized value so that larger scores indicate better performance. 
(i) $\boldsymbol{IL}$  (\emph{Inverse Latency}), a standardized delay score derived from the mean realized latency $\overline{T}_{rly}$ ;
(ii) $\boldsymbol{Trly}$ (\emph{Inverse Delay Std}), a stability score obtained by reversing the standard deviation $\mathrm{std}(T_{rly})$;
(iii) $\boldsymbol{SuS}$ (\emph{Success Rate}), the proportion of orders successfully completed (or satisfying the speedup constraint), reflecting reliability;
(iv) $\boldsymbol{Sp}$  (\emph{Speedup}), the average acceleration ratio (e.g., $T_e/T_{rly}$), reflecting efficiency improvement over the baseline processing time;
(v) $\boldsymbol{LG}$ (\emph{Load Gini Score}), a reversed Gini-based score computed from institution-level processed value, where higher scores indicate lower load imbalance;
(vi) $\boldsymbol{Fair}$ (\emph{Fairness Gini Score}), a reversed Gini-based score computed from individual utilities, where higher scores indicate better fairness among participants.
Finally, we compute an $\boldsymbol{Overall}$ score by averaging the above normalized factors, yielding a composite measure for cross-method ranking.

For \textbf{RQ2}, we investigate how different \emph{collaboration network structures} affect the service utility of the data service market under the \emph{same} order stream and identical entity settings. We treat each network topology as a structural baseline and compare the resulting performance at the \emph{individual}, \emph{organizational}, and \emph{system} levels. Specifically, we consider eight representative collaboration topologies deployed within each institution (i.e., \textbf{GovA}: Isolated Node Network; \textbf{GovB}: Globally Coupled Network; \textbf{GovC}: Star Network; \textbf{GovD}: One-Dimensional Ring; \textbf{GovE}: G(N,L) Random Network; \textbf{GovF}: WS Small-World Network; \textbf{GovG}: NW Small-World Network; \textbf{GovH}: BA Scale-Free Network). These topologies provide a controlled spectrum from no collaboration links to fully connected collaboration, as well as structured and stochastic networks with shortcuts or hubs. All other factors (order arrivals, agent costs and capabilities, and the time-generation mechanism) are fixed so that any observed differences can be attributed to network structure rather than environmental variation.
Following the multi-layer measurement framework, we evaluate the impact of network structure using quantitative metrics at three levels. At the \textbf{individual level}, we report the mean individual utility $\overline{U}_{\mathrm{ind}}$ and its dispersion $\sigma(U_{\mathrm{ind}})$ to capture both effectiveness and stability of participants' outcomes; we further compute the inequality of individual utility via the Gini coefficient $\mathrm{Gini}_i$. At the \textbf{organizational level}, we report the mean organizational utility $\overline{U}_{\mathrm{org}}$ and the relative improvement Gains(\%) with respect to the reference topology, reflecting the organization-level value creation under different collaboration patterns. At the \textbf{system level}, we report the overall system utility $\overline{U}_{\mathrm{sys}}$ together with its dispersion $\sigma(U_{\mathrm{sys}})$ to assess market-level efficiency and robustness. Finally, we provide a \textit{Rank} over topologies based on the system-level outcomes (with higher $\overline{U}_{\mathrm{sys}}$ and lower $\sigma(U_{\mathrm{sys}})$ preferred), enabling a direct comparison of how structural choices shape service utility under the same experimental conditions.

\subsection{Analysis of Experimental Results}

\subsubsection{Analysis of Service Utility Measurement Method (RQ1) }

RQ1 investigates whether the proposed system utility measurement $U_{\text{sys}}$ can evaluate the data service market more \emph{stably} and \emph{effectively} than representative alternatives. To this end, we compare $U_{\text{sys}}$ with three widely adopted baselines, including the DeLone \& McLean proxy (D\&M), the QoS-utility baseline (QoS), and the Serv-qual proxy (SaL). All methods are evaluated under \emph{identical} order streams, identical entity settings, and the same time-generation mechanism; only the \emph{measurement criterion} is changed. This protocol guarantees that any performance differences are attributable to the measurement method rather than environmental variation. The key novelty of $U_{\text{sys}}$ lies in its \emph{cross-layer consistency}: it unifies individual-, organizational-, and system-level contributions into a coherent platform-level objective, thereby producing a measurement that is simultaneously performance-sensitive, cost-aware, and stability-oriented.

\begin{figure}[htbp]
\centerline{\includegraphics[width=0.99\linewidth]{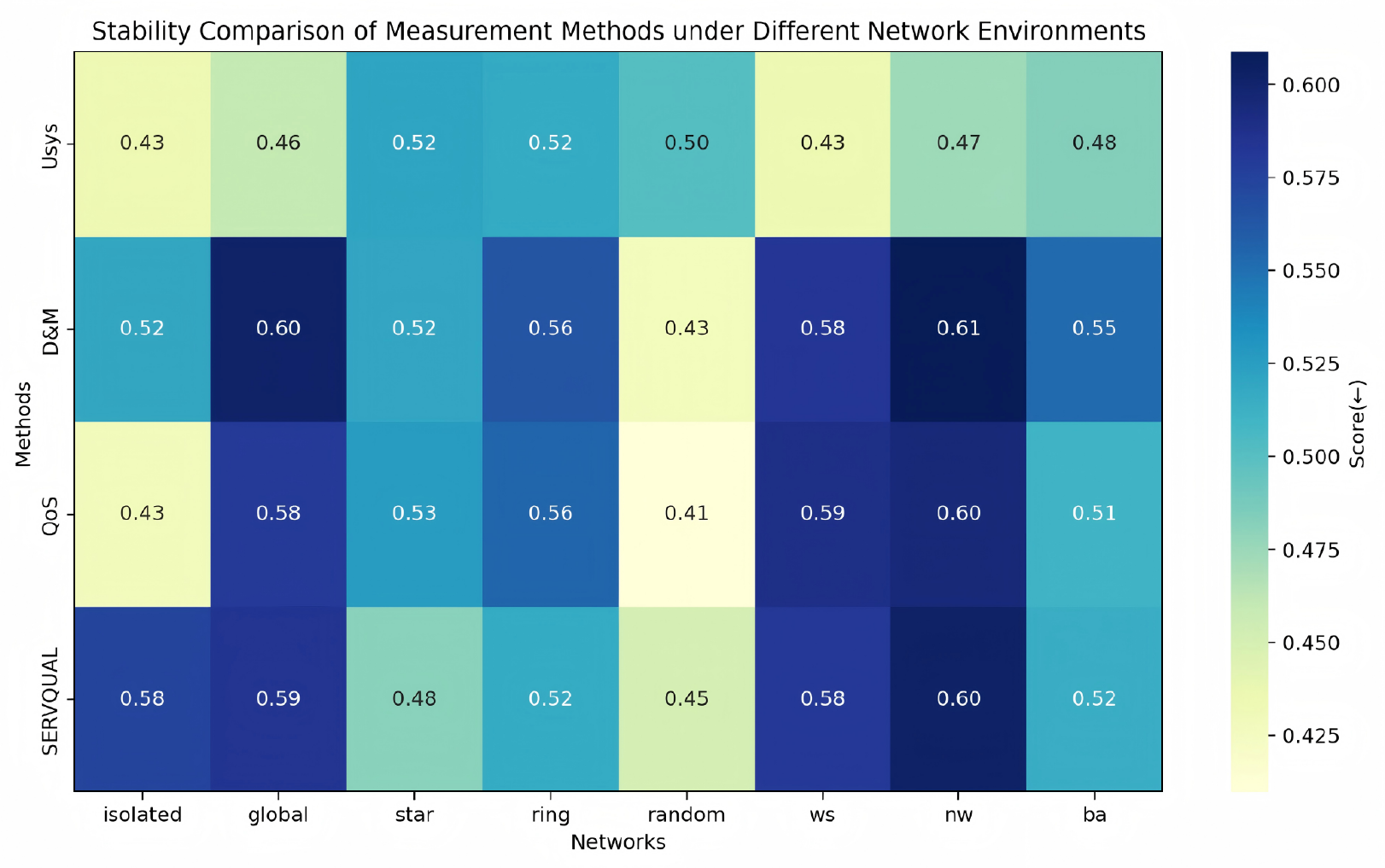}}
\caption{The Stability Comparison of Various Measurement Methods under Different Network Environments}
\vskip 10.1pt%
\label{fig:Stability Comparison}
\end{figure}

Table~\ref{tab:meas_norm_comp} reports the normalized comparison results with consistent \emph{higher-is-better} semantics. Among all compared methods, $U_{\text{sys}}$ achieves the highest Overall score (0.83), outperforming both D\&M and QoS (each $0.80$) and substantially exceeding SaL (0.69). This indicates that $U_{\text{sys}}$ provides the best balanced assessment across reliability, efficiency, and equity-related dimensions, rather than optimizing a single aspect at the expense of others. Fig.~\ref{fig:Stability Comparison} further evaluates the stability of each measurement method under different collaboration network environments. Across these heterogeneous topologies, $U_{\text{sys}}$ maintains consistently competitive stability scores without the pronounced drops observed for the baselines in certain network settings. Such robustness is crucial for data service markets where collaboration patterns and interaction constraints can vary significantly. The stability advantage aligns with our structure-aware modeling and multi-layer utility design: by embedding realized delivery time and coordination effects into utility accounting, $U_{\text{sys}}$ remains reliable even when network-induced coordination and occupancy change.
\begin{table}[H]
\centering
\caption{Comparison of Measurement Methods.}
\label{tab:meas_norm_comp}
\renewcommand{\arraystretch}{1.1}
\setlength{\tabcolsep}{4pt} 
\begin{tabular}{lccccccc}
\toprule
\textbf{Method} & \textbf{IL} & \textbf{Trly} & \textbf{SuS} & \textbf{Sp} & \textbf{LG} & \textbf{Fair} & \textbf{Overall} \\
\midrule
D\&M      & \underline{0.97} & \underline{0.95} & 0.81 & \underline{0.98} & \textbf{0.91} & 0.97 & \underline{0.80} \\
QoS       & \underline{0.97} & \underline{0.95} & 0.81 & \underline{0.98} & \underline{0.90} & \textbf{1.00} & \underline{0.80} \\
SaL  & 0.76 & 0.57 & \textbf{1.00} & 0.90 & \textbf{0.91} & 0.58 & 0.69 \\
$U_{sys}$     & \textbf{0.99} & \textbf{1.00} & \underline{0.94} & \textbf{1.00} & 0.86 & \underline{0.98} & \textbf{0.83} \\
\bottomrule
\end{tabular}
\vspace{2pt}

{\footnotesize \raggedright \hspace{2em} \textit{Note.} Best in each column is in \textbf{bold} and second is \underline{underlined}. \par}
\end{table}

A closer inspection of the factor-wise results in Table~\ref{tab:meas_norm_comp} explains why $U_{\text{sys}}$ dominates the composite score.
First, on latency-related dimensions, $U_{\text{sys}}$ attains the best performance on both $\boldsymbol{IL}$ (0.99) and $\boldsymbol{Trly}$ (1.00), implying lower mean latency and higher delivery stability. This is because $U_{\text{sys}}$ explicitly internalizes realized delivery time into the platform-level value, discouraging inefficient collaboration patterns that incur excessive coordination overhead or congestion-amplified delays.
Second, on reliability and efficiency, $U_{\text{sys}}$ achieves the best success rate $\boldsymbol{SuS}$ (0.94) and speedup $\boldsymbol{Sp}$ (1.00). In contrast, D\&M and QoS reach the same moderate $\boldsymbol{SuS}$ level (0.81), while SaL achieves the highest $\boldsymbol{SuS}$ (1.00) but does not translate this into a balanced advantage in other dimensions. This highlights that reliability alone is insufficient; an effective market-level measurement must also reward efficient and stable fulfillment.
Third, on equity-related dimensions, $U_{\text{sys}}$ maintains strong fairness with $\boldsymbol{Fair}=0.98$ while keeping a high load-balance score $\boldsymbol{LG}=0.86$. Notably, SaL performs well on $\boldsymbol{LG}$ (0.91) but poorly on $\boldsymbol{Fair}$ (0.58), suggesting that perceived-quality-oriented proxies may encourage allocations that appear balanced at the institution level but still yield inequitable outcomes among individuals. QoS attains the best fairness score ($\boldsymbol{Fair}= 1.00$) but does not dominate in stability and latency-related metrics, reflecting the limitation of purely QoS-driven aggregation when coordination and congestion effects are present.

In summary, the results confirm that the proposed $U_{\text{sys}}$ provides a more stable and effective evaluation for the data service market than representative baselines. In particular, $U_{\text{sys}}$ achieves the highest composite score and exhibits strong robustness across heterogeneous collaboration networks, validating that our multi-layer utility decomposition and structure-aware modeling yield a faithful and actionable measurement method under realistic network variability.

\subsubsection{Analysis of Network Structure Impact on Service Utility (RQ2)}
\label{sec:net_impact_utility}

\begin{figure}[htbp]
\centerline{\includegraphics[width=0.99\linewidth]{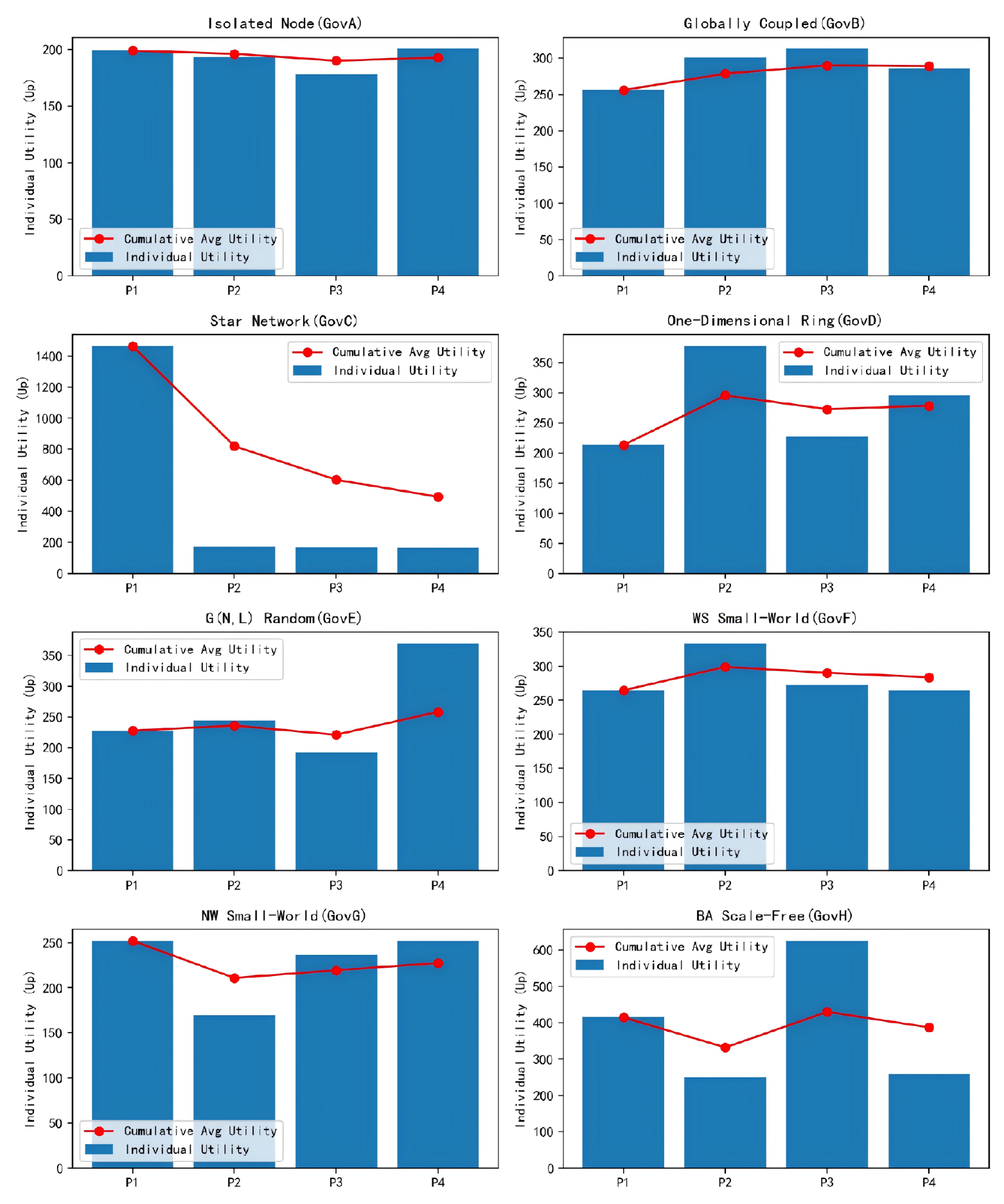}}
\caption{The Individual Utility of Salesman Agents in Different Government Agencies}
\vskip 10.1pt%
\label{fig:Individual Utility}
\end{figure}

\begin{figure}[htbp]
\centerline{\includegraphics[width=0.99\linewidth]{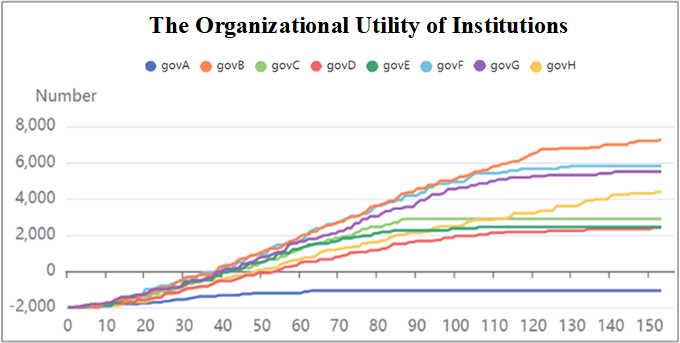}}
\caption{The Organizational Utility of Institutions in Data Service Market}
\vskip 10.1pt%
\label{fig:Organizational Utility}
\end{figure}

\begin{figure}[htbp]
\centerline{\includegraphics[width=0.99\linewidth]{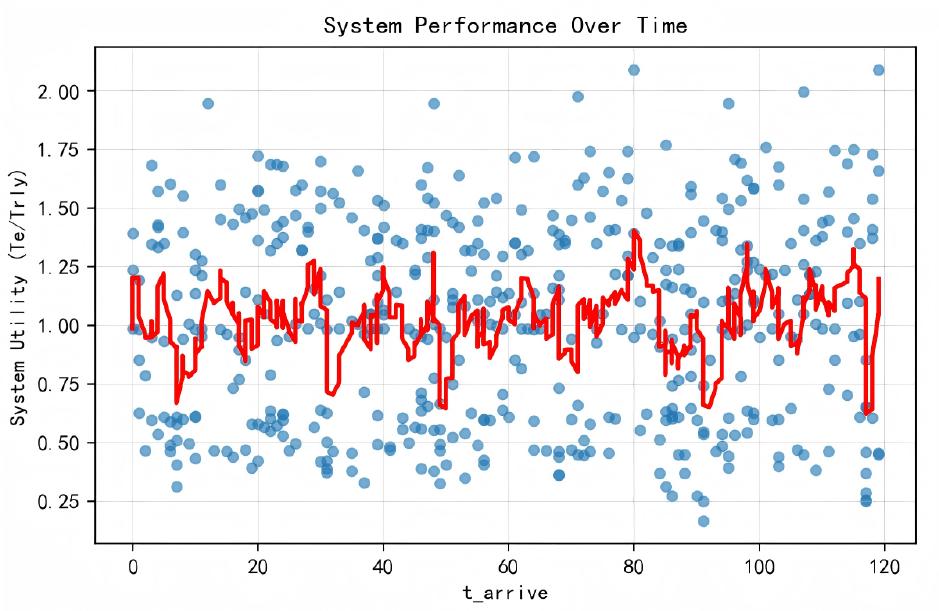}}
\caption{The System Performance Over Time}
\vskip 10.1pt%
\label{System Performance Over Time}
\end{figure}

\begin{figure}[htbp]
\centerline{\includegraphics[width=0.99\linewidth]{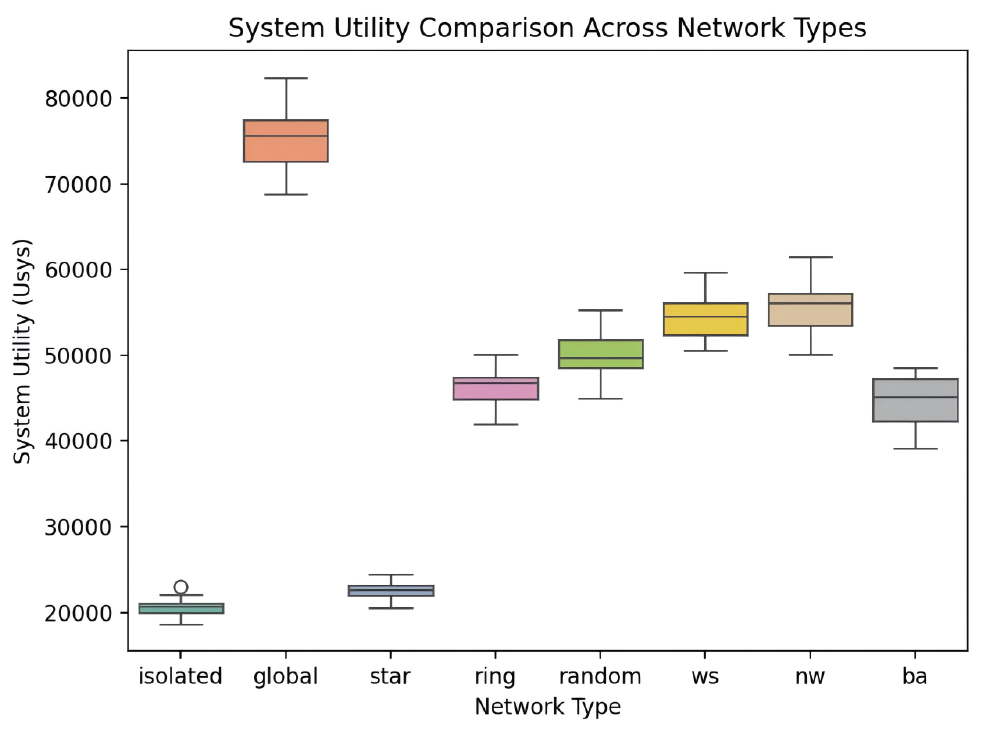}}
\caption{The System Utility with Different Networks}
\vskip 10.1pt%
\label{fig:System Utility}
\end{figure}

This subsection investigates how collaboration topologies structurally constrain coordination processes and consequently shape service utilities at the individual, organizational, and system levels. Table~\ref{tab:quant_summary_topology} reports the key statistics (mean, dispersion, Gini, and rank), while Fig.~\ref{fig:Individual Utility}-Fig.~\ref{fig:System Utility} further visualize individual utility distributions, organizational utility accumulation, temporal system dynamics, and robustness.

Individual-level utility: centralization inflates the mean but harms fairness; WS small-world is best in stability and equity.
At the individual level, GovC (Star) attains the largest mean individual utility $\bar U_{\mathrm{ind}}=492.595$, yet it exhibits severe inequality (Gini $=0.496$) and extremely high dispersion ($\sigma(U_{\mathrm{ind}})=648.359$), reflecting a hub-dominance ``rich-get-richer'' effect. Fig.~5 shows clear polarization under the star structure, implying that a high mean may be driven by a few beneficiaries rather than broad improvement.
By contrast, GovF (WS Small-World) is best in both stability and fairness, achieving the smallest $\sigma(U_{\mathrm{ind}})=10.263$ and the smallest Gini $=0.024$. GovB (Globally Coupled) is second-best in fairness (Gini $=0.040$) and keeps dispersion low ($\sigma=24.680$), highlighting how dense connectivity promotes opportunity equalization and risk diversification. These findings align with our structure-constrained collaboration modeling: topology governs not only average payoffs but also the distribution and propagation of volatility.

Organizational-level utility: globally coupled yields the highest institutional utility, followed by small-world.
Fig.~\ref{fig:Organizational Utility} and Table~\ref{tab:quant_summary_topology} show that GovB (Globally Coupled) remains the best at the organizational level with $\bar U_{\mathrm{org}}=4555.205$, corresponding to the upper envelope of the accumulation trajectories. GovF (WS Small-World) is second-best ($\bar U_{\mathrm{org}}=4240.275$), and the Gains(\%) values (7.4 for GovB and 1.2 for GovF) further support the ranking. This suggests that multi-path reachability and load balancing enhance order matching and collaboration efficiency, thus accumulating higher institutional value. Conversely, GovA (Isolated) is substantially worse ($\bar U_{\mathrm{org}}=2179.753$), confirming that missing collaborative links prevent institutions from benefiting from scale effects and complementary division of labor.

System-level utility: global coupling dominates; small-world is a robust near-optimal alternative.
As shown in Table~~\ref{tab:quant_summary_topology} and Fig.~\ref{fig:System Utility}, GovB (Globally Coupled) achieves the highest mean system utility, $\bar U_{\mathrm{sys}}=75083.970$ (Rank 1), consistent with the top ``global'' distribution in the boxplot. This indicates that dense connectivity maximizes reachability and cross-agent resource sharing, lifting the upper bound of system-wide matching efficiency.
GovF (WS Small-World) and GovG (NW Small-World) form the second tier with $\bar U_{\mathrm{sys}}=55456.279$ (Rank 2) and $54472.563$ (Rank 3), respectively. By combining local clustering with a small number of long-range shortcuts, small-world structures shorten functional paths and mitigate localized congestion without requiring full connectivity, yielding a favorable efficiency--cost trade-off.
In contrast, GovA (Isolated) is the worst ($\bar U_{\mathrm{sys}}=20528.649$, Rank 8) and GovC (Star) is also low ($22574.292$, Rank 7), suggesting that insufficient connectivity (isolated) or hub bottlenecks (star) fundamentally limit throughput and collaboration gains.

Regarding system variability $\sigma(U_{\mathrm{sys}})$ , GovC (Star) is the smallest (1047.402) and GovA (Isolated) is second (1123.526), yet both have low mean system utilities (Rank 7/8). This shows that low variance does not necessarily imply high effectiveness: star networks may operate stably at a low throughput due to centralized scheduling bottlenecks, while isolated structures cap collaboration gains by design. As shown in Fig.~\ref{System Performance Over Time}, although transient perturbations occur frequently with order arrivals, the system utility fluctuates within a bounded range and quickly reverts to its mean level, indicating short-lived shocks rather than persistent drift across network structures.Fig.~\ref{fig:System Utility} further indicates temporal responses under workload fluctuations; globally coupled and small-world networks better absorb localized congestion thanks to alternative paths, maintaining higher-utility regimes more robustly.

\begin{table*}[htbp]
\centering
\caption{Quantitative Comparison across Collaboration Topologies.}
\label{tab:quant_summary_topology}
\renewcommand{\arraystretch}{1.15}
\setlength{\tabcolsep}{6pt}
\begin{tabular}{l r r r r r r r r}
\toprule
\textbf{Topology} &
\multicolumn{3}{c}{\textbf{Individual-level}} &
\multicolumn{2}{c}{\textbf{Organizational-level}} &
\multicolumn{3}{c}{\textbf{System-level}} \\
\cmidrule(lr){2-4}\cmidrule(lr){5-6}\cmidrule(lr){7-9}
& $\bar U_{\text{ind}}\uparrow$
& $\sigma(U_{\text{ind}})\downarrow$
& Gini$\downarrow$
& $\bar U_{\text{org}}(UG)\uparrow$
& Gains(\%)$\uparrow$
& $\bar U_{\text{sys}}\uparrow$
& $\sigma(U_{\text{sys}})\downarrow$
& Rank \\
\midrule
GovA (Isolated)          & 192.788 & 33.331 & 0.047 & 2179.753 & -48.6 & 20528.649 & \underline{1123.526} & 8 \\
GovB (Globally Coupled)  & 289.183 & \underline{24.680} & \underline{0.040} & \textbf{4555.205} & \textbf{7.4} & \textbf{75083.970} & 3322.730 & \textbf{1} \\
GovC (Star)              & \textbf{492.595} & 648.359 & 0.496 & 2883.712 & -32.0 & 22574.292 & \textbf{1047.402} & 7 \\
GovD (Ring)              & 278.502 & 75.366 & 0.126 & 3593.746 & -15.2 & 46121.258 & 1988.327 & 5 \\
GovE (Random $G(N,L)$)   & 258.352 & 77.427 & 0.133 & 3460.193 & -18.4 & 50115.238 & 2935.542 & 4 \\
GovF (WS Small-World)    & 283.681 & \textbf{10.263} & \textbf{0.024} & \underline{4240.275} & \underline{1.2} & \underline{55456.279} & 2655.588 & \underline{2} \\
GovG (NW Small-World)    & 227.573 & 39.401 & 0.072 & 4211.262 & -0.7 & 54472.563 & 2489.793 & 3 \\
GovH (BA Scale-Free)     & \underline{387.233} & 176.142 & 0.207 & 3665.261 & -13.6 & 44595.925 & 2791.403 & 6 \\
\bottomrule
\end{tabular}
\vspace{2pt}

{\footnotesize \raggedright \hspace{5em} \textit{Note.} Best in each column is in \textbf{bold} and second is \underline{underlined}. \par}
\end{table*}

In Summary, we observe consistent structure--collaboration--performance regularities: (i) Globally Coupled is best at the system and organizational levels; (ii) Small-World provides a robust near-optimal alternative; (iii) Star inflates individual mean utility but severely harms fairness and stability; and (iv) Isolated suffers from limited collaboration gains. These results validate the proposed multi-level utility decomposition and structure-constrained collaboration modeling, enabling topology comparison together with explainable diagnosis of fairness, volatility, and bottleneck mechanisms for governance-oriented network design.

\section{Conclusion}
The research presented in this paper introduces an innovative approach to understanding and analyzing the complex dynamics of data service markets through the lens of service ecosystems. By developing a comprehensive modeling, measurement and network analysis method, this study addresses the challenges about measurement  posed by the intricate interactions nature of agents within these markets. In the future, the process of researching collaboration models in the data service market needs to further consider the collaboration costs of three-level entities. In addition, based on the research results of this article, further explore the impact of other factors in the data service market on the collaboration process and incorporate more complex behaviors of agents.

\section*{Acknowledgment}
This work was supported by the National Natural Science Foundation of China (No.92367202).

\bibliographystyle{IEEEtran}
\bibliography{ref}

\end{document}